# Single-shot measurement of free-electron laser polarization at SDUV-FEL

Lie Feng, Haixiao Deng*, Tong Zhang, Chao Feng, Jianhui Chen, Xingtao Wang, Taihe Lan, Lei Shen
Wenyan Zhang, Haifeng Yao, Xiaoqing Liu, Bo Liu, Dong Wang

*Shanghai Institute of Applied Physics, Chinese Academy of Sciences, Shanghai, 201800, P. R. China*

In this paper, a division-of-amplitude photopolarimeter (DOAP) for measuring the polarization state of free-electron laser (FEL) pulse is described. The incident FEL beam is divided into four separate beams, and four Stokes parameters can be measured in a single-shot. In the crossed-planar undulators experiment at Shanghai deep ultraviolet FEL test facility, this DOAP instrument constructed in house responses accurately and timely while the polarization-state of fully coherent FEL pulses are switched, which is helpful for confirming the crossed-planar undulators technique for short-wavelength FELs.

**Keywords**: polarization, DOAP, Stokes parameters, seeded-FEL

PACS numbers: 41.60.Cr

## 1. Introduction

Polarization is regarded as one of the essential attributes of light, which is utilized as information modulation carrier. Considering its significant roles in the field of physics, chemistry, biology and material sciences, high-intensity radiation pulses with well-defined and flexible polarization [1-4] are of great interest in synchrotron radiation light sources and free-electron lasers (FELs) community. Generally, due to the coupling between the radiation field and the electron motion in undulators, a planar undulator only generates linearly polarized light, while an elliptically polarized undulator (EPU) is capable of providing radiation pulses with arbitrary polarization.

Crossed-planar undulators is an alternative way to generate flexibly polarized pulse for FEL sources [5-10], which is based on the superposition of horizontal and vertical radiation fields generated from two planar undulators in a crossed configuration. A pulsed electromagnetic phase-shifter between the crossed undulators is used to slip the phase and hence to flexibly and fast switch the polarization of the combined radiation. More recently, the first experimental demonstration of polarization switching using crossed-planar undulators [11-13] has been successfully accomplished at the Shanghai deep ultraviolet free-electron laser (SDUV-FEL) facility [14-15]. In the experiment, the dependence of the polarization handedness as a function of the phase between the two crossed undulators was measured and found to be in good agreement with theory, which indicates that scaling crossed-planar undulators to seeded X-ray FELs may be possible.

In this paper, we report the single-shot polarization measurement of FEL radiations at SDUV-FEL, i.e., an in house constructed division-of-amplitude photopolarimeter (DOAP) [16-18]. In section 2, the principle of the DOAP is briefly described. Section 3 mainly concentrates on the crossed-planar undulators experiment setup and the calibration of the DOAP system. The polarization measurements of FEL pulses from crossed-planar undulators are illustrated and discussed in Section 4. This paper is concluded with the final remarks in Section 5.

## 2. Principle of DOAP

The DOAP system at SDUV-FEL mainly consists of three modules: (1) the calibration module of polarization-station generation (PSG), (2) the measurement module of polarization-state detection (PSD), and (3) the data-processing module of A/D card and computer. The schematic of the instrument is shown in Fig. 1.

The core of the DOAP system is PSD. It consists of three non-polarizing beam splitter cubes with a metallic-dielectric coating which allows the transmission and reflection for different polarization states to be within 50% of each other, four polarizers, one quarter-wave plate and four photodetectors. The incident light beam entering the PSD is divided into four beams by three beam splitters ($BS_1$, $BS_2$ and $BS_3$), and then each beam experiences different polarized elements (i.e., with different azimuth angles of four polarizers) and reaches four photodetectors $D_1$, $D_2$, $D_3$, $D_4$, whose output intensities are $I_1$, $I_2$, $I_3$, $I_4$ respectively. The detector signals are obtained and analyzed by A/D card and computer.

The measurement of four Stokes parameters can be considered by the Muller matrix algebra. Let the **S** be the Stokes parameters of the incident beam, $\mathbf{S}_R$ and $\mathbf{S}_T$ be the reflected and transmitted beams' Stokes parameters divided by every beam splitters. **S** is the unknown vector under measurement,

$$\mathbf{S} = [S_0, S_1, S_2, S_3] \tag{1}$$

$$\mathbf{S}_R = \mathbf{RS} \tag{2}$$

$$\mathbf{S}_T = \mathbf{TS} \tag{3}$$

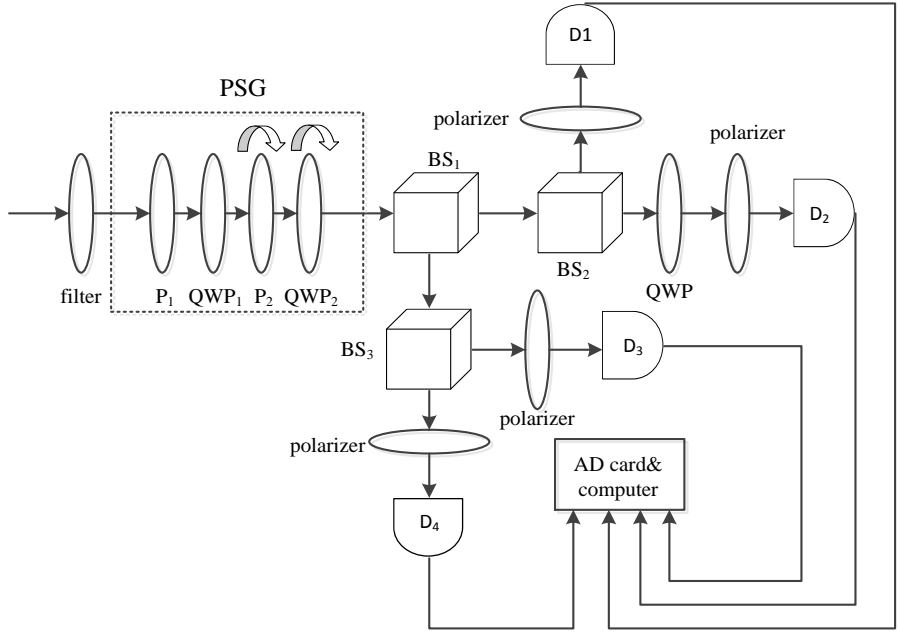

Fig.1. Schematic of the DOAP system constructed in house.

Here **R** and **T** indicates the Muller matrix of reflection and transmission by the beam splitter. If we assume $\mathbf{\Gamma}(\varphi_k)$ be the Muller matrix of the different optical component behind each beam splitters, where $k$=1, 2, 3, and 4, respectively. Thus $I_1$, $I_2$, $I_3$ and $I_4$ are given by

$$\begin{aligned}
I_1 &= C_1 \mathbf{T}_1 \mathbf{R}_2 \mathbf{\Gamma}(\varphi_1) \mathbf{S} \\
I_2 &= C_2 \mathbf{T}_1 \mathbf{T}_2 \mathbf{\Gamma}(\varphi_2 + \pi/2) \mathbf{S} \\
I_3 &= C_3 \mathbf{R}_1 \mathbf{R}_3 \mathbf{\Gamma}(\varphi_3) \mathbf{S} \\
I_4 &= C_4 \mathbf{R}_1 \mathbf{T}_3 \mathbf{\Gamma}(\varphi_4) \mathbf{S}
\end{aligned} \tag{4}$$

where $C_k$ and $\varphi_k$ is the photodetector sensitivity and the polarizer azimuth angle of the $k$-th beam, respectively. And the azimuth angle of the quarter-wave plate is set to be 0 degree. Then Eq. (4) can be simplified:

$$\mathbf{I} = \mathbf{AS} \tag{5}$$

**A** is a 4×4 matrix (i.e., **A** must be nonsingular), called instrument matrix, that is a characteristic of the DOAP system at a given wavelength and parameter setup. If **A** is known, the Stokes parameters of the incident beam can be calculated from the inverse relationship:

$$\mathbf{S} = \mathbf{A}^{-1} \mathbf{I} \tag{6}$$

For a DOAP system shown in Fig. 1, the instrument matrix **A** depends on the transmission and reflection matrix of the beam splitters, azimuth angles of all polarizers and the sensitivity of all photon detectors. However, it is not necessary to measure all these elements, because the matrix **A** can be calibrated by the system itself. If the incident beam polarized in four known and linearly independent states described by Stokes parameters $\mathbf{S}_c$, according to Eq. (5), the instrument matrix **A** can be derived entirely with the corresponding intensities $\mathbf{I}_c$ measured by photon detectors. Thus, the PSG is the polarizing optics including two pair of polarizers and quarter-wave plates for self-calibration. $P_1$ and $QWP_1$ ensure that the incident beam is fully circular polarized, while the followed $P_2$ and $QWP_2$ are mounted on the motorized rotators.

Under such circumstances, it is possible to obtain a serial of determined polarization-state for DOAP calibration with different combinations of $P_2$ and $QWP_2$ azimuths angles. Then with the calibrated instrument matrix **A**, the polarization states of the incident beam can be figured out by the readout of the photon detectors.

## 3. Experiment setup and DOAP calibration

The layout of crossed-planar undulators experiment based on the seeded FEL at SDUV-FEL is illustrated in Fig. 2. It mainly consists of an electromagnetic planar modulator (EMU65), a dispersive chicane, and a crossed-planar radiators system (PMU-H, phase-shifter and PMU-V). Firstly, the 135MeV electron beam passes through the modulator EMU65, where the beam energy is modulated by the 1047nm laser. Then the energy modulation is converted to the longitudinal density modulation in a dispersive chicane. Finally, the microbunched electron beam goes through the crossed-planar undulators, where it emits coherent 523nm radiation with horizontally and vertically linear polarization-state respectively. Therefore, using the electromagnetic phase-shifter, the output polarization-state of the combined FEL is expected to be flexibly switched pulse by pulse. The detailed parameters of the crossed-planar undulators experiment at SDUV-FEL can be found in Ref. 13.

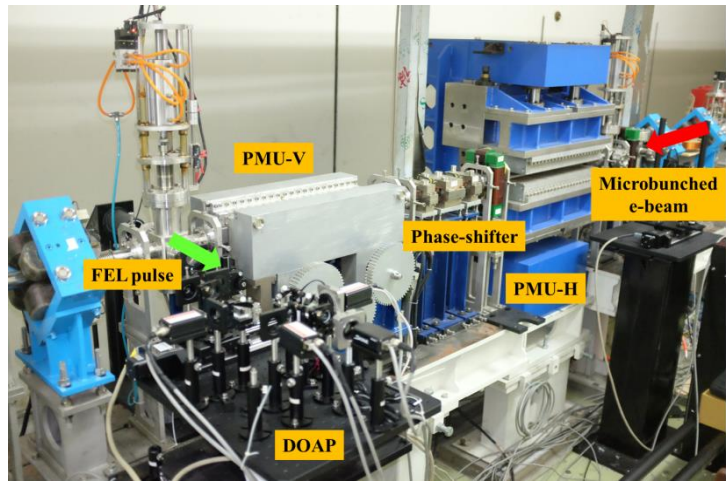

Fig. 2. Crossed-planar undulators experiment beam line at SDUV-FEL.

The 2nd harmonic of the 1047nm seed laser from Nd-YLF laser is used to calibrate the DOAP. Firstly in PSG module, by appropriate setting of $P_1$ and $QWP_1$, the laser beam is considered to be circularly polarized before $P_2$ and $QWP_2$. Then we choose the following four pairs of azimuth angles for $P_2$ and $QWP_2$, (0°, 0°), (0°, 45°), (0°, 90°) and (45°, 45°). The corresponding Stoke parameters of these four polarization-states $S_C$ is theoretically calculated, with detecting the output intensities matrix $I_c$, the DOAP instrument matrix **A** is solved according to Eq. (6).

After calibration, the PSG module is moved away and the spontaneous emission polarization from a planar undulator is measured to check the validity of DOAP system. We shut down the PMU50-H and keep the PMU50-V resonant at 523nm, and the spontaneous emission from PMU-V is sent to the DOAP system, which is considered to be vertically linear polarized. With the calibrated instrument matrix **A**, Fig. 3 shows the measured polarization degree of the 523nm spontaneous radiation in 100 continuous shots. The mean linear polarization degree and the standard deviation of the undulator spontaneous emission is about 96.4% and 11.5%, which is well consistent with the undulator radiation theory and confirms the validity of the DOAP instrument.

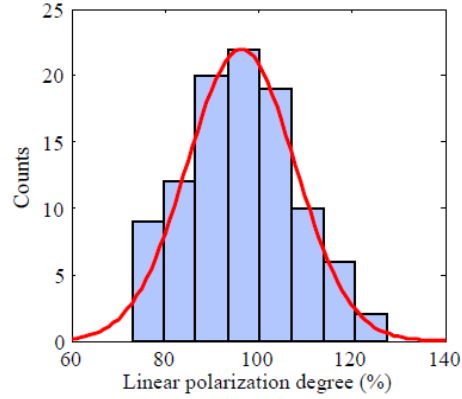

Fig. 3. Linear polarization degree statistics of undulator spontaneous emission from 100 shots measurement.

## 4. Crossed-planar undulators experiment results

In the crossed-planar undulators experiment at SDUV-FEL, the spontaneous emission is firstly investigated. We turn off the 1047nm seed laser and change the magnetic gap of PMU-H from maximum to 35mm (523nm resonance) while the PMU-V is resonant on 523nm, the measured results of combined spontaneous emission is shown in Fig. 4. When the PMU-H is fully opened, the spontaneous radiation is mainly emitted from the PMU-V, which is fully vertical polarized. With the gap decreased, the horizontal field from PMU-H is enhanced and the polarized degree of combined spontaneous emission decreases. Hence, the curve indicates that the linear polarization degree changes from 100% to 0%. This result illustrates that the combined 523nm spontaneous emission from the crossed-planar undulators is nearly non-polarized, which depends on the fact of random phases of the vertical and horizontal spontaneous emission.

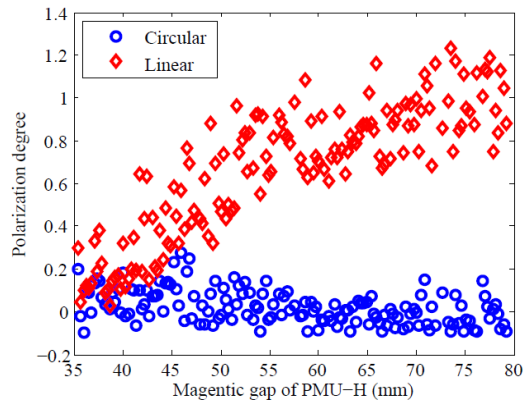

Fig. 4. Measured polarization of the combined spontaneous emission of crossed-planar undulators

Then we turn on the seed laser and try to measure the polarization-state and polarized degree of the combined coherent 523nm FEL radiation from the cross-planar undulators at SDUV-FEL. With the optimized experimental conditions and a relatively stable beam, the relationship between the polarization degrees and the slip phase of the phase-shifter are plotted in Fig. 5. With the circular polarization degree is raised from 10% to 80%, while the linear polarization degree (45) is reduced from 90% to 10%. It successfully demonstrates the ability of polarization switching between linear polarization and circular polarization by using crossed-planar undulators. Moreover, the switching between the linear and circular polarization can be repeated reasonably when further increasing the phase-shifter strength. It is worth stressing that, the large polarization fluctuations in the measurement is mainly contributed by the beam instability at SDUV-FEL due to the lack of a feedback system and the time jitter of the seed laser.

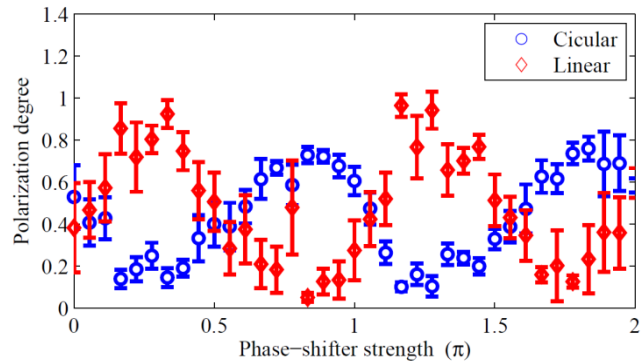

Fig. 5. Polarization switching of coherent 523nm FEL pulse emitted from crossed-planar undulators, with tuning of the phase-shifter, measured by the DOAP system.

## 5. Conclusions

In this paper, a DOAP system is used to characterize the qualitative polarization-state and quantificational polarization degree of the seeded FEL from cross-planar undulators, which is the first time to measure FEL polarization at SDUV-FEL test facility. During the measurement, the DOAP instrument constructed in house responses accurately and timely. Our results indicate that polarization of the combined FEL pulse can be flexibly switched if the superposition of the vertical and horizontal component of coherent emissions is perfect in spatial and temporal, which confirms the physics behind the crossed-planar undulators. It is expected that this technique can be extended to the existing [1] and forthcoming seeded FEL facilities [19-21] from EUV to soft X-ray spectral region, with the help of the most recently advanced seeding schemes [22-24].

## Acknowledgement

The authors are grateful to all the SDUV-FEL colleagues for helpful discussions, experiment setup establishment and commissioning assistance. This work was partially supported by the Major State Basic Research Development Program of China (2011CB808300) and the National Natural Science Foundation of China (11175240, 11205234 and 11322550).